\documentclass[twocolumn,superscriptaddress,amssymb,amsmath,nobibnotes,aps,prlgroupedaddress]{revtex4}  % for review and submission
\usepackage{graphicx}  % needed for figures
\usepackage{dcolumn}   % needed for some tables
\usepackage{bm}        % for math
\usepackage{amssymb}   % for math
\usepackage[]{amsmath,amssymb}
\usepackage{graphics,epsfig}
\usepackage{float}
\usepackage{mathtools}
\usepackage{esint}

\begin{document}

%\input author_list.tex       % D0 authors (remove the first 3 lines
                             % of this file prior to submission, they
                             % contain a time stamp for the authorlist)
                             % (includes institutions and visitors)
%\date{\today}

\title{Modular Hamiltonian of a chiral fermion on the torus}

\author{David Blanco}
\email{dblanco@df.uba.ar}
\affiliation{Departamento de F\'{\i}sica, FCEN, Universidad de Buenos Aires and IFIBA-CONICET\\
1428 Buenos Aires, Argentina}
\author{Guillem P\'erez-Nadal}
\email{guillem@df.uba.ar}
\affiliation{Departamento de F\'{\i}sica, FCEN, Universidad de Buenos Aires and IFIBA-CONICET\\
1428 Buenos Aires, Argentina}

\date{\today}

\begin{abstract}
We consider a chiral fermion at non-zero temperature on a circle (i.e., on a torus in the Euclidean formalism) and compute the modular Hamiltonian corresponding to a subregion of the circle. We do this by a very simple procedure based on the method of images, which is presumably generalizable to other situations. Our result is non-local even for a single interval, and even for Neveu-Schwarz boundary conditions. To the best of our knowledge, there are no previous examples of a modular Hamiltonian with this behavior.
\end{abstract}

\maketitle

\section{Introduction}

\noindent In recent years, the study of entanglement and its measures has proven to be very useful in unveiling some of the deepest properties of Quantum Field Theory (QFT). Entanglement measures are based on the reduced density matrix, or equivalently on (minus) its logarithm, the modular Hamiltonian. 
Among other applications,
the knowledge of modular Hamiltonians was essential for the proof of the averaged null energy condition \cite{Faulkner:2016mzt}, the derivation of quantum energy inequalities \cite{Blanco:2013lea,Blanco:2017akw} and the formulation of a well-defined version of the Bekenstein bound \cite{Casini:2008cr}. Modular Hamiltonians also played a key role in applications to holography, the most notable case probably being the derivation of the linearized Einstein equations in the bulk from entanglement properties of the boundary Conformal Field Theory (CFT) \cite{Faulkner:2013ica,Lashkari:2013koa,Blanco:2018riw,Swingle:2014uza}.

There are only few cases where modular Hamiltonians have been computed. The result is universal and local for the vacuum of any QFT reduced to Rindler space \cite{Unruh:1976db,Bisognano:1976za}, and from this result one can also derive CFT expressions for the 
vacuum reduced to a ball in the plane \cite{Casini:2011kv}, for a thermal state reduced to an interval in the plane in 1+1 dimensions \cite{Hartman:2015apr} and for the vacuum reduced to an interval in the cylinder in 1+1 dimensions \cite{Cardy:2016fqc}. In these cases the modular Hamiltonian turns out to be local, but non-local contributions are expected to appear in general. This was first shown explicitly with the calculation of the modular Hamiltonian of the vacuum state reduced to an arbitrary set of disjoint intervals in 1+1 dimensions for free chiral fermions on the plane \cite{Casini:2009vk}, later for the cylinder \cite{Klich:2015ina} and more recently for free chiral scalars on the plane \cite{Arias:2018tmw}. Another notable result is the modular Hamiltonian for the vacuum state of any QFT reduced to regions ending on a null plane \cite{Casini:2017roe}.

In this paper we compute a new modular Hamiltonian, namely that corresponding to a chiral fermion on the circle at non-zero temperature (i.e., on the torus in Euclidean language). 
Our analysis is based on the method of images applied to the calculation of the Euclidean propagator, which enables us to map the problem to a similar problem on the plane.
The method turns out to be very simple, and we expect it to have applications beyond the case of chiral fermions. Our result is non-local even for a single interval, even for Neveu-Schwarz (antiperiodic) boundary conditions. 

\section{Modular Hamiltonian from the resolvent}

\noindent Consider a chiral fermion $\psi$ on a circle of length $L$. The Hamiltonian is
\begin{equation}\label{hamil}
    H=\pm i\int_{-L/2}^{L/2}dx\,\psi^\dagger\psi'\,,
\end{equation}
where the sign depends on the chirality. Suppose that the field is in a thermal state with inverse temperature $\beta$. The purpose of this paper is to compute the reduced density matrix $\rho_V$ corresponding to a subset $V$ of the circle or, equivalently, the modular Hamiltonian
\begin{equation}
    H_V=-\log \rho_V.
\end{equation}
Since the global state is Gaussian, the reduced density matrix is also Gaussian and hence the modular Hamiltonian has the form
\begin{equation}
    H_{V}=\int_V dx dy\,\psi^{\dagger}(x)K_{V}(x,y)\psi(y).
\end{equation}
As shown in \cite{peschel2003calculation}, the kernel $K_{V}$ is related to the
two-point function $G_V(x,y)=\langle\psi(x)\psi^{\dagger}(y)\rangle$ ($x,y\in V$) by 
\begin{equation}\label{H2}
    K_{V}=-\log\left(G_{V}^{-1}-1\right),
\end{equation}
where both $K_V$ and $G_V$ are viewed as operators acting on functions on $V$. 
This equation can be rewritten as
\begin{equation}\label{Hresol}
    K_V=-\int_{1/2}^\infty d\xi\left[R_V(\xi)+R_V(-\xi)\right],
\end{equation}
where $R_V$ is the resolvent of $G_V$,
\begin{equation}
    R_V(\xi)=\frac{1}{G_V+\xi-1/2},
\end{equation}
as can be easily checked by explicitly performing the integral in (\ref{Hresol}). Thus, the problem of computing the modular Hamiltonian reduces to that of finding the resolvent of $G_V$.

\section{The method of images}

\noindent Our strategy for computing the resolvent is based on the method of images applied to the calculation of the Euclidean propagator $G$, which is defined by
\begin{alignat}{2}\label{eucl}
G(x,t;y,u)&=
    \theta(t-u)
    \langle e^{H(t-u)}\psi(x)e^{-H(t-u)}\psi^\dagger(y)\rangle\nonumber\\ 
    &-\theta(u-t)\langle \psi^\dagger(y)e^{H(t-u)}\psi(x)e^{-H(t-u)}\rangle 
\end{alignat}
for $t-u\in(-\beta,\beta)$
and by analytic continuation for other values of $t$ and $u$, where $\theta$ is the step function. 
Depending on the spin structure chosen on the circle, the Euclidean propagator can be either periodic or antiperiodic in $x$ with period $L$, and it is 
antiperiodic in $t$ with period $\beta$. Due to these quasiperiodicity properties, we may view $G$ as a section of a line bundle over a torus of circumferences $L$ and $\beta$. Inserting (\ref{hamil}) in (\ref{eucl}) one sees that the Euclidean propagator satisfies
\begin{equation}\label{diffeucl}
    (\partial_t\pm i\partial_x)G=\delta(x-y)\delta(t-u)
\end{equation}
for $(x,t)-(y,u)\in(-L,L)\times(-\beta,\beta)$. Identifying ${\mathbb R}^2$ with ${\mathbb C}$ via the map $(x,t)\mapsto x\mp it$, this equation says precisely that $G(z,w)$ is analytic in $z$ for 
$z\ne w$ and has a simple pole at $z=w$ with residue $\pm 1/(2\pi i)$. In other words,
\begin{equation}\label{eucleq}
    G(z,w)=\pm\frac{1}{2\pi i}\frac{1}{z-w}+F(z,w),
\end{equation}
where $F$ is analytic in $z$ for $z-w\in(-L,L)\times(-\beta,\beta)$. In the language of complex variables the quasiperiodicity conditions read
\begin{equation}\label{bc}
    G(z+P_i,w)=(-1)^{\nu_i}G(z,w),
    %G(z+L,w)=(-1)^s G(z,w)\qquad G(z+i\beta,w)=-G(z,w).
\end{equation}
where $P_1=L$, $P_2=i\beta$, $\nu_1\in\{0,1\}$ and $\nu_2=1$.
Eqs.~(\ref{eucleq}) and (\ref{bc}) have a unique solution. Indeed, the difference $\Delta G$ between two solutions
is analytic for $z-w\in(-L,L)\times(-\beta,\beta)$ and satisfies (\ref{bc}), so it is analytic and bounded throughout the complex plane. By Liouville's theorem, such a function is necessarily a constant, so the antiperiodicity in the imaginary direction implies $\Delta G=0$. In order to find the solution, let us first look at the limiting case $L,\beta\to\infty$, where the torus becomes a plane and the quasiperiodicity conditions (\ref{bc}) are replaced by the condition that $G$ vanish at infinity. Since the only analytic function that vanishes at infinity is the zero function, the solution of Eq.~(\ref{eucleq}) on the plane is
\begin{equation}\label{propplane}
    G(z,w)=\pm\frac{1}{2\pi i}\frac{1}{z-w}\equiv G_0(z,w).
\end{equation}
Going back to the torus, i.e., to generic values of $L$ and $\beta$, we can solve Eqs.~(\ref{eucleq}) and (\ref{bc}) by the method of images,
\begin{equation}\label{images0}
    G(z,w)=\sum_{\lambda\in\Lambda}(-1)^{\nu\cdot\lambda}G_0(z+\lambda,w),
\end{equation}
where $\Lambda$ is the lattice
\begin{equation}
    \Lambda=\{\lambda_1 P_1+\lambda_2 P_2,\, \lambda_i\in{\mathbb Z}\}.
\end{equation}
and $\nu\cdot\lambda=\nu_1\lambda_1+\nu_2\lambda_2$.
Indeed, the function (\ref{images0}) clearly has the form (\ref{eucleq}), and one can easily check that it satisfies the quasiperiodicity conditions (\ref{bc}).

\begin{figure}[t]
    \centering
    \includegraphics[width=8.6cm]{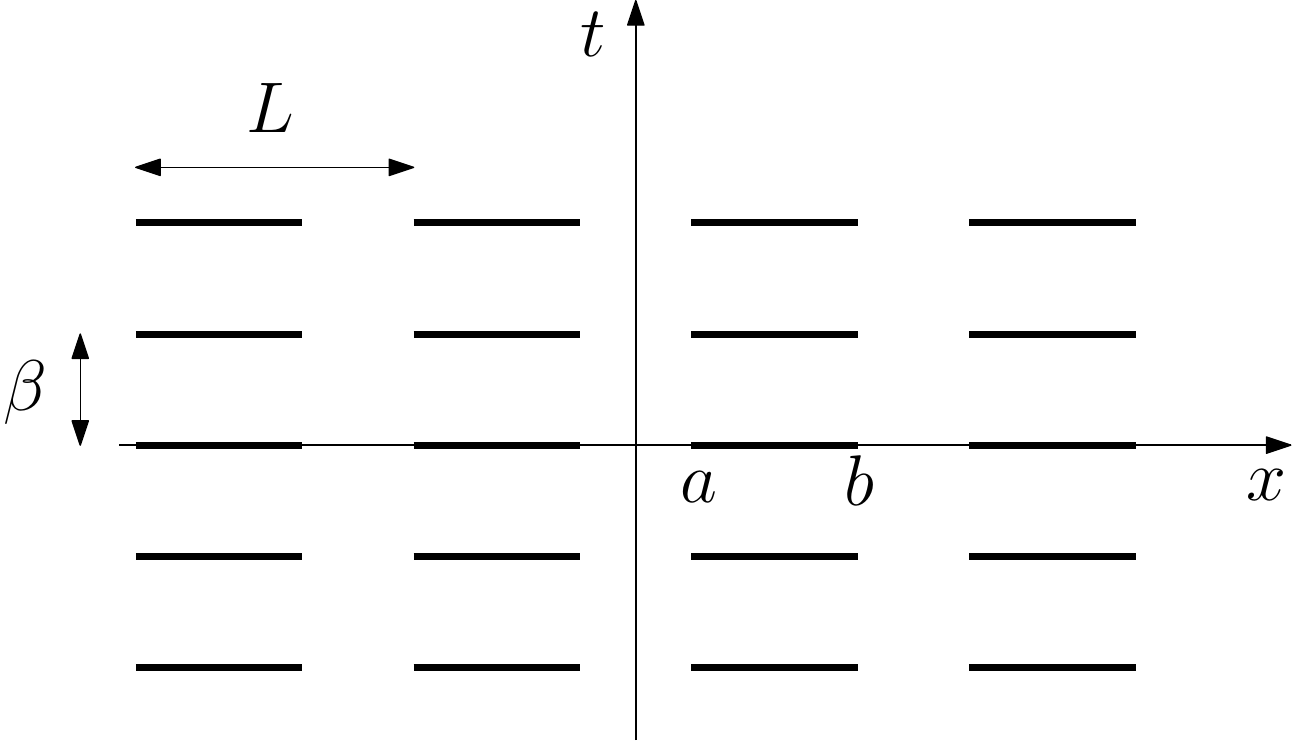}
    \caption{The region $V_\Lambda$ when $V$ is the interval $(a,b)$.}
    \label{fig1}
\end{figure}

Let now $x,y\in V$. It is clear from (\ref{eucl}) that $G_V(x,y)=G(x,0^+;y,0)$ (note that it is important to take the limit $t\to 0$ from above, because if we take it from below we pick a delta function). With our identification of ${\mathbb R}^2$ with ${\mathbb C}$, we thus have
\begin{equation}\label{2pointprop}
    G_V(x,y)=G(x\mp i\epsilon,y),
\end{equation}
so, by (\ref{images0}),
\begin{equation}\label{images}
    G_V(x,y)=\sum_{\lambda\in{\Lambda}}(-1)^{\nu\cdot\lambda} G_{0V_\Lambda}(x+\lambda,y),
\end{equation}
where $V_\Lambda=\bigcup_{\lambda\in\Lambda}(V+\lambda)$, see Fig.~\ref{fig1},
and
\begin{equation}
    G_{0V_\Lambda}(u,v)=G_0(u\mp i\epsilon,v)
\end{equation}
for $u,v\in V_\Lambda$. The main reason why the method of images is useful for us is that Eq.~(\ref{images}) also holds for the powers of the operators involved,
\begin{equation}\label{imagesp}
     (G_{V}^n)(x,y)=\sum_{\lambda\in\Lambda}(-1)^{\nu\cdot\lambda}(G_{0V_\Lambda}^{n})(x+\lambda,y)
\end{equation}
for any $n\in{\mathbb N}$, where $(A^n)(u,v)$ denotes the kernel of the operator $A^n$ (not to be confused with $[A(u,v)]^n$). We can see this by induction. First, the above equation is satisfied for $n=1$ (this is Eq.~(\ref{images})). And second, if it holds for some $n\in{\mathbb N}$ we have

{\footnotesize{
\begin{alignat}{2}
&(G_{V}^{n+1})(x,y) =\int_V dz\, G_{V}(x,z)\,(G_{V}^{n})(z,y)\nonumber\\
&=\sum_{\lambda,\mu\in\Lambda}(-1)^{\nu\cdot(\lambda+\mu)}\int_V dz\, G_{0V_\Lambda}(x+\lambda,z)\,( G_{0V_\Lambda}^n)(z+\mu,y)\nonumber\\
&= \sum_{\lambda,\mu\in\Lambda}(-1)^{\nu\cdot(\lambda+\mu)}\int_V dz\, G_{0V_\Lambda}(x+\lambda+\mu,z+\mu)\,(G_{0V_\Lambda}^n)(z+\mu,y)\nonumber\\
&= \sum_{\lambda',\mu\in\Lambda}(-1)^{\nu\cdot\lambda'}\int_{V+\mu}dz'\, G_{0V_\Lambda}(x+\lambda',z')\,( G_{0V_\Lambda}^{n})(z',y)\nonumber\\
&= \sum_{\lambda'\in\Lambda}(-1)^{\nu\cdot\lambda'}\int_{V_\Lambda}dz'\, G_{0V_\Lambda}(x+\lambda',z')\,( G_{0V_\Lambda}^n)(z',y)\nonumber\\
&=\sum_{\lambda'\in\Lambda}(-1)^{\nu\cdot\lambda'}(G_{0V_\Lambda}^{n+1})(x+\lambda',y).
\end{alignat}
}}

\noindent In the third equality we have used the translational invariance of $G_0$, and in the fourth we have defined $\lambda'=\lambda+\mu$ and $z'=z+\mu$. Eq.~(\ref{imagesp}) implies that the method of images works for any function of $G_V$ which can be expressed as a power series. In particular, it works for the resolvent,
\begin{equation}\label{resimages}
    R_{V}(\xi;x,y)=\sum_{\lambda\in\Lambda}(-1)^{\nu\cdot\lambda}R_{0V_\Lambda}(\xi;x+\lambda,y).
\end{equation}
In the case of zero temperature, $\beta\to\infty$, the terms with $\lambda_2\ne 0$ do not contribute to the sum (\ref{images0}), so the lattice $\Lambda$ effectively reduces to $\{mL,m\in{\mathbb Z}\}$ and, in consequence, the region $V_\Lambda$ reduces to an arrangement of segments in the real line. The resolvent $R_{0V_\Lambda}$ is well-known in that case \cite{Casini:2009vk}, so we can use it to obtain $R_V$ via the above equation. To the best of our knowledge, $R_{0V_\Lambda}$ is not known for generic temperatures, where $V_\Lambda$ is a collection of segments distributed all over the complex plane, but it can be easily computed as we will explain in the next section.

\section{The resolvent for a generic set of segments in the plane}

\begin{figure}
\centering
\includegraphics[width=8.6cm]{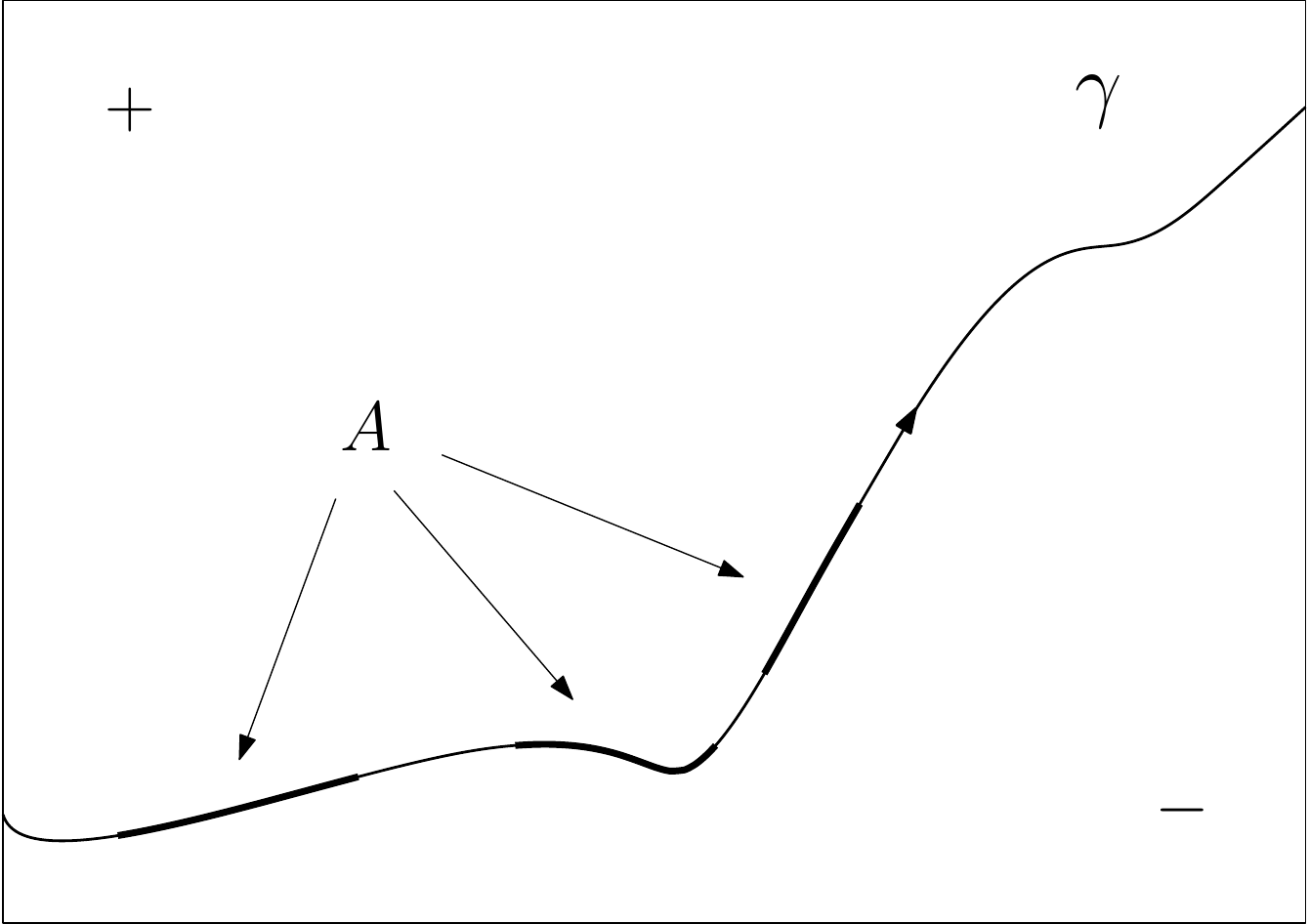}
\caption{
A curve $\gamma$ in the complex plane, and a subset $A\subset\gamma$. The $+/-$ region is the region to the left/right of the curve.
}
\label{fig2}
\end{figure}

\noindent Consider a curve $\gamma$ in the complex plane with both endpoints at infinity, and a subset $A\subset\gamma$, see Fig.~\ref{fig2}. As shown in the figure, $\gamma$ divides the plane into two regions: the one to the left of the curve ($+$ region) and the one to the right ($-$ region); if $\gamma$ is the real line these regions are the upper and lower half-planes respectively.
The purpose of this section is to compute the resolvent of the operator $G_{0A}$ with kernel
\begin{equation}\label{Cv}
    G_{0A}(u,v)=G_0(u^\mp,v)=\pm\frac{1}{2\pi i}\frac{1}{u^\mp-v}
\end{equation}
for $u,v\in A$, where $F(u^\mp)$ denotes the limit of $F$ as $u$ is approached from the $\mp$ region. This resolvent is known in the case where $\gamma$ is the real line \cite{Casini:2009vk}; as we will see, the computation for $\gamma$ generic is remarkably simple.

We will first obtain an expression for the powers of $G_{0A}$. Then we will insert that expression into the expansion of the resolvent in powers of $G_{0A}$ and find that it is easy to perform the sum. For the square we have
\begin{alignat}{2}\label{squareker}
    (G_{0A}^2)(u,v)&=\int_A dw \,G_{0A}(u,w)G_{0A}(w,v)\nonumber\\
    &=\frac{1}{(2\pi i)^2}\int_A \frac{dw}{(u^\mp-w)(w-v^\pm)}\nonumber\\
   &=\frac{1}{2\pi i}G_{0A}(u,v)\left[\omega_{A}^\pm(u)+\omega_{A}^\mp(v)\right],
\end{alignat}
where
\begin{equation}\label{omega}
    \omega_{A}^\pm(u)=\pm\int_A \frac{dv}{u^\mp-v}.
\end{equation}
Note that
\begin{equation}\label{sumomega}
    \omega_{A}^++\omega_{A}^-=2\pi i.
\end{equation}
Indeed, if $A^\pm$ is a slight deformation of $A$ which has the same endpoints but travels through the $\pm$ region we have
\begin{alignat}{2}
    \omega_{A}^+(u)+\omega_{A}^-(u)&=\int_A \frac{dv}{u^--v}-\int_A \frac{dv}{u^+-v}\nonumber\\
    &=\int_{A^+} \frac{dv}{u-v}-\int_{A^-} \frac{dv}{u-v}\nonumber\\
    &=\ointclockwise \frac{dv}{u-v}=2\pi i,
\end{alignat}
where the contour in the last integral encircles $A$, and hence $u$. We can rewrite (\ref{squareker}) as an operator equation,
\begin{equation}\label{square}
    G_{0A}^2=\frac{1}{2\pi i}\left(\omega_{A}^\pm G_{0A}+G_{0A}\omega_{A}^\mp\right).
\end{equation}
Now, using (\ref{sumomega}) and (\ref{square}) it is a simple matter to check that the operator-valued function
\begin{equation}\label{barG}
    \bar G_{0A}(s)=(1+s)^{-\omega_{A}^\pm/(2\pi i)}\,G_{0A}\,(1+s)^{-\omega_{A}^\mp/(2\pi i)}
\end{equation}
satisfies $\bar G_{0A}'=-\bar G_{0A}^2$. In turn, this implies for the $n$-th derivative $\bar G_{0A}^{(n)}=(-1)^nn!\bar G_{0A}^{n+1}$, as can be easily shown by induction. Noting that $\bar G_{0A}(0)=G_{0A}$, we thus obtain
\begin{equation}
    G_{0A}^{n+1}=\frac{(-1)^n}{n!}\bar G_{0A}^{(n)}(0).
\end{equation}
Inserting this expression into the expansion of the resolvent in powers of $G_{0A}$ (which is a geometric series) we recognize the Taylor series of $\bar G_{0A}$,
\begin{alignat}{2}\label{resol1}
    R_{0A}(\xi)&=\frac{1}{G_{0A}+\xi-1/2}\nonumber\\
    &=\frac{1}{\xi-1/2}\left[1-\frac{1}{\xi-1/2}\sum_{n=0}^\infty\frac{(-1)^n}{(\xi-1/2)^n}G_{0A}^{n+1}\right]\nonumber\\
    &=\frac{1}{\xi-1/2}\left[1-\frac{\bar G_{0A}(1/(\xi-1/2))}{\xi-1/2}\right]\nonumber\\
    &=\frac{1}{\xi-1/2}\left[1-\frac{e^{-ik(\xi)\omega_{A}^\pm}G_{0A}e^{ik(\xi)\omega_{A}^\pm}}{\xi+1/2}\right],
\end{alignat}
where in the last step we have used (\ref{sumomega}) and defined
\begin{equation}
    k(\xi)=\frac{1}{2\pi}\log\frac{\xi-1/2}{\xi+1/2}.
\end{equation}
Eq.~(\ref{resol1}) gives the resolvent for a generic subset $A$ of a generic curve. Let us particularize it to the case where $A$ is a collection of horizontal segments, $A=\bigcup_\alpha\left[(a_\alpha,b_\alpha)+i\eta_\alpha\right]$ with $a_\alpha,b_\alpha,\eta_\alpha\in{\mathbb R}$. In this case the integral (\ref{omega}) is easily computed,
\begin{alignat}{2}\label{omega2}
    \omega_{A}^\pm(u)=\pm\sum_\alpha\log\frac{a_\alpha+i\eta_\alpha-u\pm i\epsilon}{b_\alpha+i\eta_\alpha-u\pm i\epsilon},
\end{alignat}
and, using the relation $1/(x\mp i\epsilon)=1/x\pm i\pi\delta(x)$ (there is a principal part implicit in the first term), the resolvent (\ref{resol1}) takes the form
\begin{alignat}{2}\label{resol2}
    R_{0A}(\xi;u,v)=\frac{1}{\xi^2-1/4}&\Bigg\{\xi\delta(u-v)\nonumber\\
    &\mp \frac{e^{-ik(\xi)[\omega_{A}^\pm(u)-\omega_{A}^\pm(v)]}}{2\pi i(u-v)}\Bigg\}.
\end{alignat}
This result agrees with that of \cite{Casini:2009vk} in the case where $A$ is contained in the real line.

\section{Modular Hamiltonian on the torus}

\noindent Our last step is to use the resolvent just computed to obtain the resolvent on the torus by the method of images, Eq.~(\ref{resimages}), and from it the modular Hamiltonian. For simplicity, we concentrate on the case where $V$ is a single interval, $V=(a,b)$,
but the analysis that follows extends straightforwardly to the general case of multiple intervals. Setting $A=V_\Lambda$ in (\ref{omega2}) yields
\begin{equation}
    \omega_{V_\Lambda}^\pm(u)=\pm\sum_{\lambda\in\Lambda}\log\frac{a-u+\lambda\pm i\epsilon}{b-u+\lambda\pm i\epsilon}.
\end{equation}
This series is ambiguous because it is not absolutely convergent. However, its second derivative is unambiguous,
\begin{alignat}{2}
    (\omega_{V_\Lambda}^\pm)''(u)&=\mp\sum_{\lambda\in\Lambda}\left[\frac{1}{(a-u+\lambda\pm i\epsilon)^2}-\frac{1}{(b-u+\lambda\pm i\epsilon)^2}\right]\nonumber\\
    &=\mp\left[\wp(a-u\pm i\epsilon)-\wp(b-u\pm i\epsilon)\right],
\end{alignat}
where $\wp$ is the Weierstrass elliptic function (see \cite{sigma} for a review). Since the latter is related to the Weierstrass sigma function
\begin{equation}
    \sigma(z)=z\prod_{\lambda\ne 0}\left(1+\frac{z}{\lambda}\right)e^{-\frac{z}{\lambda}+\frac{1}{2}\left(\frac{z}{\lambda}\right)^2}
\end{equation}
by $\wp=-(\log\sigma)''$, we conclude that
\begin{equation}\label{omegaLambda}
    \omega_{V_\Lambda}^\pm(u)=\omega^\pm(u)+c^\pm u+d^\pm,
\end{equation}
where 
\begin{equation}\label{omegapm}
    \omega^\pm(u)=\pm\log\frac{\sigma(a-u\pm i\epsilon)}{\sigma(b-u\pm i\epsilon)}
\end{equation}
and $c^\pm,d^\pm$ are undetermined constants. 
The Weierstrass sigma function is quasiperiodic,
\begin{equation}\label{sigmap}
    \sigma(z+\lambda)=(-1)^{\lambda_1+\lambda_2+\lambda_1\lambda_2}e^{\lambda\cdot\zeta(P/2)(2z+\lambda)}\sigma(z),
\end{equation}
where $\zeta=(\log\sigma)'$ and $\lambda\cdot\zeta(P/2)=\lambda_1\zeta(P_1/2)+\lambda_2\zeta(P_2/2)$. Therefore, $\omega^\pm$ is also quasiperiodic,
\begin{equation}\label{omegaLambdap}
    \omega^\pm(u+\lambda)=\omega^\pm(u)\pm2l\lambda\cdot\zeta(P/2),
\end{equation}
where $l=b-a$.
Substituting the resolvent (\ref{resol2}) with $A=V_\Lambda$ into (\ref{resimages}), and using (\ref{omegaLambda}), (\ref{omegapm}) and (\ref{omegaLambdap}), we obtain for the resolvent on the torus
\begin{alignat}{2}\label{resoltorus}
    R_V(\xi;x,y)=\frac{1}{\xi^2-1/4}&\Bigg[\xi\delta(x-y)\nonumber\\
    -&\frac{e^{\mp ik(\xi)\Delta\omega(x,y)}}{2\pi i}F^\pm(\xi;x,y)\Bigg],
\end{alignat}
where
\begin{equation}
    \Delta\omega(x,y)=\log\frac{\sigma(a-x)\sigma(b-y)}{\sigma(b-x)\sigma(a-y)},
\end{equation}
which does not need the regulator $\epsilon$ because the argument of the logarithm is real and positive, 
and
\begin{alignat}{2}
    F^\pm(\xi;z,w)=\pm\sum_{\lambda\in\Lambda}&(-1)^{\nu\cdot\lambda}e^{\mp 2ik(\xi)l\lambda\cdot\zeta(P/2)}\nonumber\\
    &\times \frac{e^{-i k(\xi)c^\pm(z+\lambda-w)}}{z+\lambda-w}.
\end{alignat}
Note that the constant $d^\pm$ has dropped out because $R_{0V_\Lambda}$ only involves the difference $\omega_{V_\Lambda}^\pm(u)-\omega_{V_\Lambda}^\pm(v)$, see (\ref{resol2}). Note also that the summand above may diverge exponentially as $\lambda$ grows; preventing this divergence fixes $c^\pm=\mp 2l\zeta(P_2/2)/P_2$. Therefore, the ambiguity in $\omega^{\pm}_{V_\Lambda}$ disappears from the resolvent on the torus. Now, $F^\pm$ has the following properties: (i) it is analytic in $z$ for $z-w\in(-L,L)\times(-\beta,\beta)$ except for a simple pole at $z=w$ with residue $\pm 1$; and (ii) it is quasiperiodic,
\begin{alignat}{2}\label{Fp}
    F^\pm(\xi;z+P_i,w)=(-1)^{\nu_i}e^{\pm 2ik(\xi)l\zeta(P_i/2)}F^\pm(\xi;z,w).
\end{alignat}
By the argument we gave under Eq.~(\ref{bc}), there is only one function with these properties. This function is
\begin{equation}\label{F}
    F^\pm(\xi;z,w)=\pm\frac{1}{\sigma(z-w)}\frac{\sigma_\nu(z-w\pm ik(\xi)l)}{\sigma_\nu(\pm ik(\xi)l)},
\end{equation}
where
\begin{equation}
    \sigma_\nu(z)=e^{-[\nu_1\zeta(P_2/2)+\zeta(P_1/2)]z}\sigma(z+\nu_1P_2/2+P_1/2)
\end{equation}
(recall that $\nu_2=1$).
Indeed, $\sigma$ is analytic, has a simple zero at the origin with $\sigma'(0)=1$ and does not vanish anywhere else in the region $(-L,L)\times(-\beta,\beta)$. Together with quasiperiodicity, this implies that the second ratio in (\ref{F}) is analytic, from which property (i) follows. On the other hand, Eq.~(\ref{sigmap}) and the relation $P_2\zeta(P_1/2)-P_1\zeta(P_2)=i\pi$ imply
\begin{equation}\label{sigmanup}
    \sigma_\nu(z+P_i)=(-1)^{\nu_i+1}e^{\zeta(P_i/2)(2z+P_i)}\sigma_\nu(z),
\end{equation}
from which property (ii) follows. Thus, Eqs.~(\ref{resoltorus}) and (\ref{F}) give the resolvent on the torus. Inserting it into (\ref{Hresol}), noting that the terms with a delta function cancel and changing the variable of integration from $\xi$ to $k(\xi)$ yields the modular Hamiltonian,
\begin{alignat}{2}\label{modhamdef}
    &K_V(x,y)=\frac{\mp i}{\sigma(x-y)}\int_{-\infty}^\infty  dk\, f(k;x,y)\nonumber\\
    &f(k;x,y)=e^{-ik\Delta\omega(x,y)}\frac{\sigma_\nu(x-y+ ikl)}{\sigma_\nu(ikl)}.
\end{alignat}
We have checked that this result coincides with known results in the limits $L\to\infty$ \cite{Hartman:2015apr} and $\beta\to\infty$ \cite{Klich:2015ina}, including the presence of a non-local term in the case $\nu_1=0$ (periodic, or Ramond, boundary conditions). In order to obtain a more explicit expression for $L$ and $\beta$ generic, note from (\ref{sigmanup}) that $f$ is quasiperiodic in $k$, 
\begin{alignat}{2}\label{fp}
    f(k+\beta/l;x,y)=&e^{-i\left\{\frac{\beta}{l}\Delta\omega(x,y)-2\eta(x-y)\right\}} f(k;x,y),
\end{alignat}
where $\eta=-i\zeta(i\beta/2)$ ($\zeta$ is odd and satisfies $\zeta(z^*)=\zeta^*(z)$, so $\eta$ is real). Therefore,
\begin{equation}\label{modhamp}
    \int_{-\infty}^\infty  dk\, f(k;x,y)=I(x,y)\int_{-\frac{\beta}{2l}}^{\frac{\beta}{2l}}  dk\, f(k;x,y),
\end{equation}
where
\begin{alignat}{2}\label{I}
&I(x,y)=\sum_{n\in{\mathbb Z}}e^{-i n\left\{\frac{\beta}{l}\Delta\omega(x,y)-2\eta(x-y)\right\}}\nonumber\\
&=2\pi\sum_{n\in{\mathbb Z}}\delta\left(\frac{\beta}{l}\Delta\omega(x,y)-2\eta(x-y)+2\pi n\right).
\end{alignat}

\begin{figure}[t]
\centering
\includegraphics[width=8.6cm]{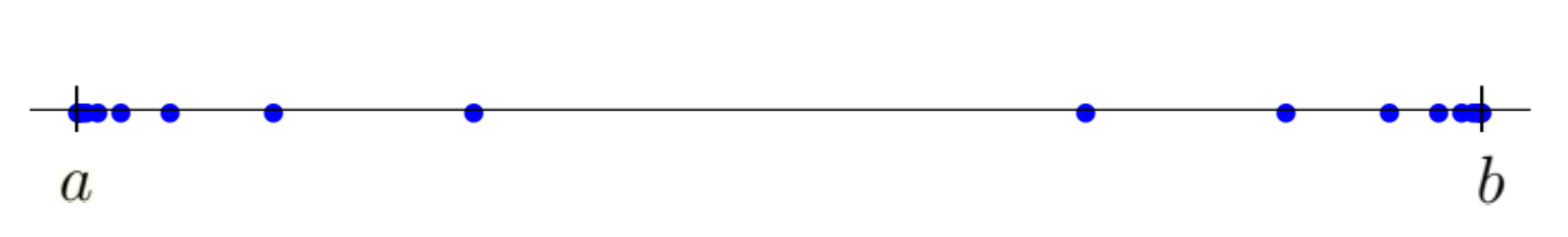}
\caption{The first few points that the non-local term (\ref{modhamfinal}) of the modular Hamiltonian couples to the midpoint of the interval, for $l/L=0.6$ and $l/\beta=0.12$. They tend to accumulate near the endpoints.}
\label{fig3}
\end{figure}

Now, for $x$ fixed, the argument of the delta function above decreases monotonically from $\infty$ to $-\infty$ as $y$ goes from $a$ to $b$, and hence it has a unique zero for each $n$. For $n=0$ this is $y=x$. Noting that $f(k;x,x)=1$ one easily obtains the local contribution to the modular Hamiltonian,
\begin{equation}
    K_{V}^{{\text{loc}}}(x,y)=\pm i\left[\tilde\beta(x)\delta'(x-y)+\frac{1}{2}\tilde\beta'(x)\delta(x-y)\right],
\end{equation}
where
\begin{equation}
    \tilde\beta(x)=\frac{2\pi}{\zeta(b-x)-\zeta(a-x)-2\eta l/\beta}.
\end{equation}
In order to obtain the non-local contribution (i.e., the contribution from the remaining values of $n$), note that the delta function in (\ref{I}) selects precisely those pairs of points $(x,y)$ for which $f$ is periodic. On the other hand, $f$ satisfies a second quasiperiodicity property, analogous to (\ref{fp}) with $\beta$ replaced by $iL$ and $\eta$ replaced by $i\zeta(L/2)$. Using these two facts, one can rewrite the integral on the right-hand side of (\ref{modhamp}) in terms of a contour integral encircling a single pole. Evaluating the latter by residues finally yields 
the non-local contribution to the modular Hamiltonian,
\begin{alignat}{2}\label{modhamfinal}
    &K_{V}^{{\text{non-loc}}}(x,y)=\frac{\pm 2i\pi^2}{l\sinh\left(\frac{L}{2l}\Delta\omega(x,y)-\zeta(L/2)(x-y)\right)}\nonumber\\
    &\times\sum_{n\ne 0}(-1)^{\nu_1 n}\delta\left(\frac{\beta}{l}\Delta\omega(x,y)-2\eta(x-y)+2\pi n\right).
\end{alignat}
For $x$ fixed, this non-local term has support in an infinite number of points which tend to accumulate near the endpoints of the interval, as shown in Fig.~\ref{fig3}. Thus, the modular Hamiltonian
is highly non-local, both for Ramond ($\nu_1=0)$ and Neveu-Schwarz ($\nu_1=1$) boundary conditions.

The above is a novel explicit example of a modular Hamiltonian, which has the interesting property of being highly non-local even for a single interval, for any choice of boundary conditions. It thus opens a new window for the exploration of the entanglement properties of QFT. While this manuscript was nearing completion a work with related results appeared \cite{Hollands:2019hje}, and another similar study \cite{Fries:2019ozf} appeared after this paper was first announced on arXiv. We leave the comparison between the three analyses for future work.

{\sl Acknowledgements.---} The authors would like to thank Horacio Casini, Pascal Fries, Alan Garbarz, Gaston Giribet, Andr\'es Goya, Mart\'in Mereb and Ignacio Reyes for very fruitful discussions. This work was supported by CONICET and Universidad de Buenos Aires, Argentina.

%%%%
\bibliography{references}

\begin{thebibliography}{10}

\bibitem{Faulkner:2016mzt}
T.~Faulkner, R.~G. Leigh, O.~Parrikar, and H.~Wang, ``{Modular Hamiltonians for
  Deformed Half-Spaces and the Averaged Null Energy Condition},'' {\em JHEP},
  vol.~09, p.~038, 2016.

\bibitem{Blanco:2013lea}
D.~D. Blanco and H.~Casini, ``{Localization of Negative Energy and the
  Bekenstein Bound},'' {\em Phys. Rev. Lett.}, vol.~111, no.~22, p.~221601,
  2013.

\bibitem{Blanco:2017akw}
D.~Blanco, H.~Casini, M.~Leston, and F.~Rosso, ``{Modular energy inequalities
  from relative entropy},'' {\em JHEP}, vol.~01, p.~154, 2018.

\bibitem{Casini:2008cr}
H.~Casini, ``{Relative entropy and the Bekenstein bound},'' {\em Class. Quant.
  Grav.}, vol.~25, p.~205021, 2008.

\bibitem{Faulkner:2013ica}
T.~Faulkner, M.~Guica, T.~Hartman, R.~C. Myers, and M.~Van~Raamsdonk,
  ``{Gravitation from Entanglement in Holographic CFTs},'' {\em JHEP}, vol.~03,
  p.~051, 2014.

\bibitem{Lashkari:2013koa}
N.~Lashkari, M.~B. McDermott, and M.~Van~Raamsdonk, ``{Gravitational dynamics
  from entanglement 'thermodynamics'},'' {\em JHEP}, vol.~04, p.~195, 2014.

\bibitem{Blanco:2018riw}
D.~Blanco, M.~Leston, and G.~P\'erez-Nadal, ``{Gravity from entanglement for
  boundary subregions},'' {\em JHEP}, vol.~18, p.~130, 2018.

\bibitem{Swingle:2014uza}
B.~Swingle and M.~Van~Raamsdonk, ``{Universality of Gravity from
  Entanglement},'' {\em arXiv e-print}, 2014.

\bibitem{Unruh:1976db}
W.~G. Unruh, ``{Notes on black hole evaporation},'' {\em Phys. Rev.}, vol.~D14,
  p.~870, 1976.

\bibitem{Bisognano:1976za}
J.~J. Bisognano and E.~H. Wichmann, ``{On the Duality Condition for Quantum
  Fields},'' {\em J. Math. Phys.}, vol.~17, pp.~303--321, 1976.

\bibitem{Casini:2011kv}
H.~Casini, M.~Huerta, and R.~C. Myers, ``{Towards a derivation of holographic
  entanglement entropy},'' {\em JHEP}, vol.~05, p.~036, 2011.

\bibitem{Hartman:2015apr}
T.~Hartman and N.~Afkhami-Jeddi, ``{Speed Limits for Entanglement},'' {\em
  arXiv e-print}, Dec 2015.

\bibitem{Cardy:2016fqc}
J.~Cardy and E.~Tonni, ``{Entanglement hamiltonians in two-dimensional
  conformal field theory},'' {\em J. Stat. Mech.}, vol.~1612, no.~12,
  p.~123103, 2016.

\bibitem{Casini:2009vk}
H.~Casini and M.~Huerta, ``{Reduced density matrix and internal dynamics for
  multicomponent regions},'' {\em Class. Quant. Grav.}, vol.~26, p.~185005,
  2009.

\bibitem{Klich:2015ina}
I.~Klich, D.~Vaman, and G.~Wong, ``{Entanglement Hamiltonians for chiral
  fermions with zero modes},'' {\em Phys. Rev. Lett.}, vol.~119, no.~12,
  p.~120401, 2017.

\bibitem{Arias:2018tmw}
R.~E. Arias, H.~Casini, M.~Huerta, and D.~Pontello, ``{Entropy and modular
  Hamiltonian for a free chiral scalar in two intervals},'' {\em Phys. Rev. D},
  vol.~98, no.~12, p.~125008, 2018.

\bibitem{Casini:2017roe}
H.~Casini, E.~Teste, and G.~Torroba, ``{Modular Hamiltonians on the null plane
  and the Markov property of the vacuum state},'' {\em J. Phys. A: Math. Gen.},
  vol.~50, no.~36, p.~364001, 2017.

\bibitem{peschel2003calculation}
I.~Peschel, ``Calculation of reduced density matrices from correlation
  functions,'' {\em Journal of Physics A: Mathematical and General}, vol.~36,
  no.~14, p.~L205, 2003.

\bibitem{sigma}
K.~Chandrasekharan, {\em {The zeta-function and the sigma-function of
  Weierstrass. In: ``Elliptic Functions. Grundlehren der mathematischen
  Wissenschaften'', vol. 281}}.
\newblock A Series of Comprehensive Studies in Mathematics, Springer, Berlin,
  Heidelberg, 1985.

\bibitem{Hollands:2019hje}
S.~Hollands, ``On the modular operator of mutli-component regions in chiral
  cft,'' {\em arXiv e-print}, Apr 2019.

\bibitem{Fries:2019ozf}
P.~Fries and I.~A. Reyes, ``{The entanglement spectrum of chiral fermions on
  the torus},'' {\em arXiv e-print}, 2019.

\end{thebibliography}
\bibliographystyle{ieeetr}
%%%%

\end{document}